\documentclass{article}

\usepackage[english]{babel}

\usepackage[letterpaper,top=2cm,bottom=2cm,left=3cm,right=3cm,marginparwidth=1.75cm]{geometry}

\usepackage{amsmath}
\usepackage{graphicx}
\usepackage[colorlinks=true, allcolors=blue]{hyperref}
\usepackage{geometry}
\usepackage{pdflscape}
\usepackage{multirow}
\usepackage{aas_macros}

\title{SAINT (Small Aperture Imaging Network Telescope) -- \\a  wide-field telescope complex for detecting and studying optical transients at times from milliseconds to years. }

\author{
Grigory Beskin$^{1,2}$, Anton Biryukov$^{6,7,2}$, Alexey Gutaev$^{1,2}$, Sergey Karpov$^{3}$,\\ Gor Oganesyan$^{4,5}$ Gennady Valyavin$^{1}$,  Azamat Valeev$^{1}$, \\Valery Vlasyuk$^{1}$, Nadezhda Lyapsina$^{1}$, Vyacheslav Sasyuk$^{2}$}

\date{Accepted to {\it Photonics} 01 December 2023}

\begin{document}

\maketitle
\noindent$^{1}$ \quad Special Astrophysical Observatory, Russian Academy of Sciences, Nizhniy Arkhyz, Russia. \\ 
$^{2}$ \quad Engelhardt Observatory, Kazan Federal University (KFU), Kazan, Russia.\\
$^{3}$ \quad CEICO, Institute of Physics, Czech Academy of Sciences, Prague, Czech Republic. \\ 
$^{4}$ \quad Gran Sasso Science Institute, L’Aquila, Italy.\\
$^{5}$ \quad INFN–Laboratori Nazionali
del Gran Sasso, L’Aquila, Italy.\\ 
$^{6}$ \quad Sternberg Astronomical Institute, Moscow State University, Moscow, Russia. \\ 
$^{7}$ \quad Faculty of Physics, HSE University, Moscow, Russia.

\begin{abstract}
In this paper, we present a project of multi-channel wide-field optical sky monitoring system with high temporal resolution -- Small Aperture Imaging Network Telescope (SAINT) --  mostly built from off-the-shelf components and aimed towards searching and studying optical transient phenomena on the shortest time scales. The instrument consists of twelve channels each containing 30cm (F/1.5) GENON Max objectives mounted on separate ASA DDM100 mounts with pointing speeds up to 50deg/s. Each channel is equipped with a 4128x4104 pixel Andor Balor sCMOS detector, and a set of photometric $griz$ filters and linear polarizers. 
At the heart of every channel is a custom built reducer-collimator module allowing rapid switching of an effective focal length of the telescope -- due to it the system is capable to operate in either wide-field survey or narrow-field follow-up modes. In the first case, the field of view of the instrument is 470 square degrees (39 sq.deg. for a single channel) and the detection limits (5$\sigma$ level at 5500\AA) are 12.5, 16.5, 19, 21  with exposure times of 20 ms, 1 s, 30 s and 20 minutes, correspondingly. 
In the second, follow-up (e.g. upon detection of a transient of interest by either a real-time detection pipeline, or upon receiving an external trigger) regime, all telescopes are oriented towards the single target, and SAINT becomes an equivalent to a monolitic 1-meter telescope, with the field of view reduced to 11$'$ x 11$'$, and the exposure times decreased down to 0.6 ms (1684 frames per second). Different channels may then have different filters installed, thus allowing a detailed study -- acquiring both color and polarization information -- of a target object with highest possible temporal resolution.
The telescopes are located in two pavilions with sliding roofs, and are controlled by a cluster of 25 computers that both govern their operation, acquire and store up to 800 terabytes of data every night, also performing its real-time processing using a dedicated fast image subtraction pipeline. Long term storage of the data will require a petabyte class storage.
The operation of SAINT will allow acquiring an unprecedented amount of data on various classes of astrophysical phenomena, from near-Earth to extragalactic ones, while its multi-channel design and the use of commercially available components allows easy expansion of its scale, and thus performance and detection capabilities.
\end{abstract}

\section{Introduction}
\begin{figure}[t]
\centering
\centerline{
    \resizebox*{1\columnwidth}{!}{\includegraphics{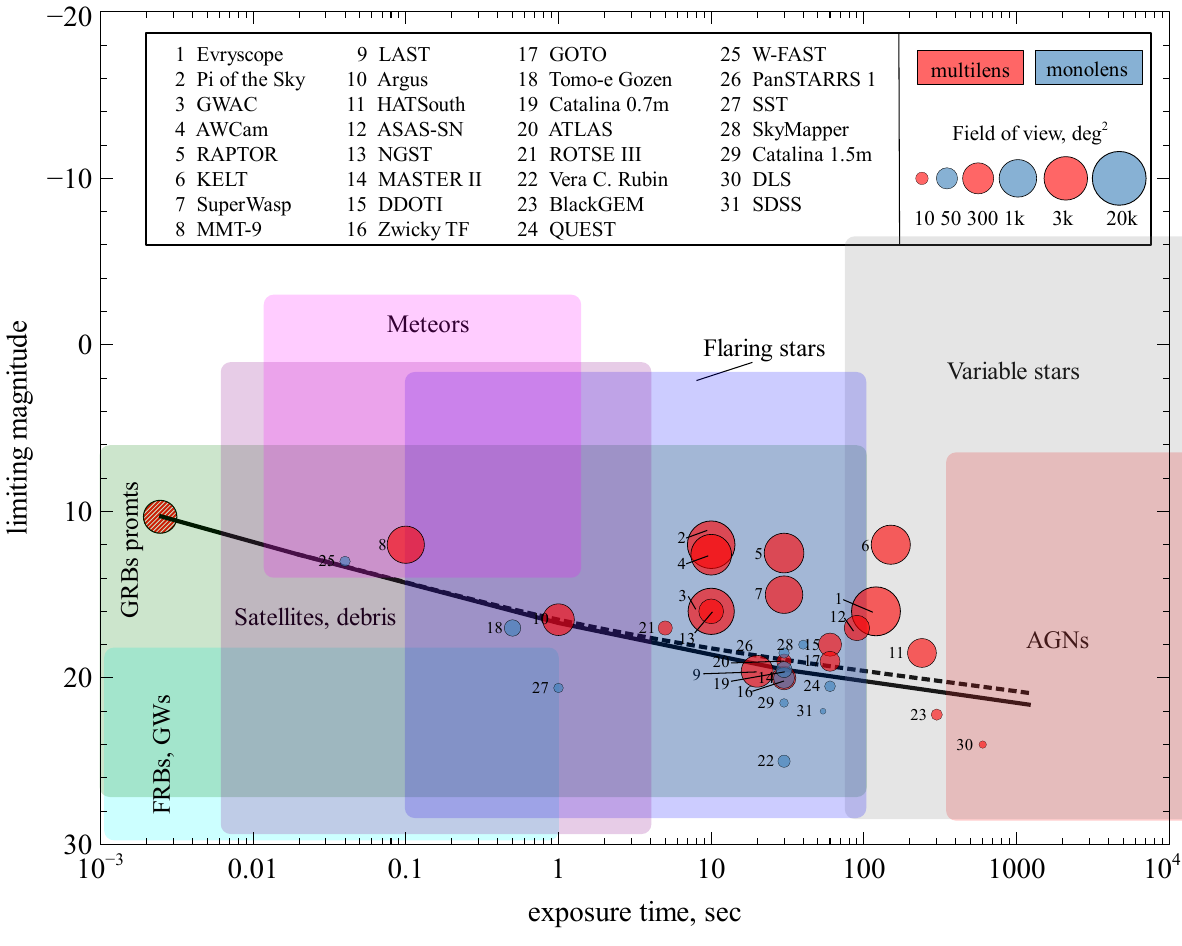}}
}
\caption{Exposure times and detection limits of various sky surveys mentioned in the text and Table~\ref{tab:surveys}. Black solid and dashed lines represent theoretical detection limit for the SAINT complex in follow-up and monitoring regimes respectively. Shaded circle represent the full field of view of SAINT.
\label{fig:exposure_limits}}
\end{figure}  

A special place among the studied astronomical phenomena belongs to non-stationary ones. In a sense, all astronomy is the science of change (evolution) of the Universe with time, both as a whole and its parts of various scales, from meteors to clusters of galaxies. It is no coincidence that in recent years a new direction has been formed and is rapidly developing - astronomy in the time domain (''Time domain astronomy''). It combines methods, tools and ideas focused on the study of non- stationary phenomena in the Universe at different time and space scales. 
The website of the IAU working group\footnote{URL: \url{http://timedomainastronomy.net/resources.html}, Master list of Time Domain Surveys section.} provides information on 62 instruments that investigate (or had investigated) variable objects. 
As long as the list of telescopes is, so is the range of specific tasks for which these instruments are intended. This includes observations of meteors, comets and asteroids, artificial satellites of the Earth, space debris, searches for the effects of microlensing and transits of exoplanets, the study of variable stars, studies of optical afterglows of gamma-ray bursts and searches for optical flares synchronous with these bursts, searches for supernovae, and the study of brightness variations of galactic nuclei, search and characterisation of optical counterparts of fast radio bursts and gravitational wave signals. The latter direction is all the more important and exceptionally interesting in connection with the discovery of the gravitational wave event GW170817 \cite{2017PhRvL.119p1101A}, caused by the merger of two neutron stars and the associated short gamma-ray burst GRB 170817A \cite{2017ApJ...848L..13A, 2017ApJ...848L..14G,Integral}, the kilonova emission\cite{kilonova_pian,kilonova_tanvir} and subsequently with a long multi-wavelength afterglow \cite{2017ApJ...848L..12A,xray_170817_1,xray_170817_2,xray_170817_3,optical_170817,radio_170817}. At the same time, there are considerations that precursors of short gamma-ray bursts can precede gravitational-wave pulses \cite{2021Galax...9..104W}, thereby marking the region of their localization, which gives hope for searching for electromagnetic radiation synchronously with the pulse itself. When searching for variability, as a rule, two modes of observation are implemented (we note a certain conventionality of such a division): survey and monitoring. In the first case, the radiation of relatively large areas of the sky is periodically recorded, while the ratio between the time of their exposure (determining the depth of the survey) and the size of the fraction of the celestial sphere observed during the night is determined by the nature of the required variability. Some of these projects and tools are SDSS \cite{2000AJ....120.1579Y}; Sky Mapper \cite{2007PASA...24....1K}; QUEST-La Silla survey \cite{2021AAS...23732403C}; ASAS-SN \cite{2017PASP..129j4502K}; Catalina Surveys \cite{2009ApJ...696..870D}; ZTF \cite{2019PASP..131a8002B}, LSST under construction \cite{2019ApJ...873..111I}. 
In the second observation mode, we are talking about the longest possible registration of the radiation of a certain object in anticipation of changes in its characteristics (intensity, spectrum, polarization, etc.). An outstanding example of this kind of TAOS II program is the search for trans-Neptunian planets by their occultations of 10,000 stars, whose long-term simultaneous monitoring of high temporal resolution (exposure - 50 ms) in individual subapertures is carried out using three telescopes with a diameter of 1.3 m \cite{2021RMxAC..53..137C}. Another classic example is the monitoring of UV Cet-type stars to study their flare activity using the largest telescopes. In particular, subsecond polarized spikes of synchrotron origin were first detected during a powerful UV Cet flare with the 6-meter telescope of SAO RAS \cite{2017PASA...34...10B}. In essence, both modes consist of monitoring, spatio-temporal in the first case, and temporal in the second.

A special place among the non-stationary phenomena studied within the framework of Time Domain Astronomy is occupied by unexpectedly appearing (and also suddenly disappearing) objects, whose localization in space and/or time is not known in advance, and whose duration is rather short - the so-called transients (transient sources). They manifest themselves as stochastic fluctuations of electromagnetic radiation of different frequencies (from radio to gamma-rays), low (tens of MeV) and high (TeV-PeV) energy neutrino events, cosmic rays, gravitational waves. 
It should be emphasized that despite the variety of forms of energy release in these events, the possibilities of understanding their nature and constructing their models are determined to a certain extent by the detection and study of their manifestations in the optical range — optical transients. These include non-stationary phenomena in the near-Earth space and the Earth's (exoplanet's?) atmosphere, non-moving phenomena such as auroras, transient luminous events (TLE) - elves, sprites, jets \cite{2015ExA....40..239A}, and moving phenomena - comets, asteroids, meteors, satellites, space debris. Finally, undoubtedly, such transients can be optical flashes (including periodic) of artificial origin - signals (transmissions) of extraterrestrial civilizations \cite{1961Natur.190..205S, 2004ApJ...613.1270H}. The characteristic duration of these phenomena range from minutes-hours to weeks-months (transits of exoplanets, outbursts of novae, supernovae, variable stars, microlensing effects) and from milliseconds to tens of seconds (gamma and radio bursts, gravitational waves, TLE, flyby meteors and satellites). In essence, we are talking about two different types of transients - long and short - the detection and study of which require the use of different methods and tools.  At the same time, within the framework of the prevailing ideas and conceptual apparatus, events of the first type are often referred to as fast transients. For example, one of the most extensive modern programs 

``Wider Faster Deeper'' \cite{2019IAUS..339..135A}, which unites about 40 ground and space instruments and is dedicated to the study of all of the fastest transients, uses the follow-up mode in the optical range (repointing of the instruments to the area of the already detected fast transients). With the use of the 4-meter Blanco telescope and DECam, at an exposure of 20 seconds, information on the subsecond fast transients is completely lost, but their environment and slow transients are successfully studied - gamma-ray burst afterglows, localization regions of fast radio bursts and gravitational wave events, bursts of novae, supernovae, red dwarfs \cite{2022AJ....163...95S}.
 
There is real contradiction between the need to use instruments with an extremely high temporal resolution in the search and study of fast optical transients and the real characteristics of telescopes, both existing and under construction.
Figure~\ref{fig:exposure_limits} and Table~\ref{tab:surveys} from Appendix demonstrate this perfectly.  We present the characteristics of 30 survey instruments focused on the search and study of non-stationary objects of various types, whose fields of view exceed 4 square degrees. 

The minimum exposure duration exceeds 10 seconds for the vast majority of these telescopes, which, given the possibility of summing up a sequence of individual frames, fully provides a depth sufficient for detecting and studying variability up to magnitudes 19–25. At the same time, these limits for systems with close fields of view may differ significantly due to the difference in the diameters of the mirrors.
Note that most of the telescopes in our list are multi-element systems. Indeed, to implement wide-angle spatial-temporal monitoring with a single aperture, a combination of initially mutually contradictory conditions is necessary - a sufficiently high detection limit (large diameter of the supply optics) and a wide field of view (short focus). This circumstance significantly limits the set of possible optical schemes of instruments that determine the optimum of such a combination. So, with an aperture diameter of D > 1–2 meters, the size of the field of view is 2–3 degrees, and with D < 1 it can reach 5–10 degrees\cite{2010amos.confE..41A, 2011AN....332..714T}. This implies the need to move to multi-telescope systems with fields of view of hundreds and thousands of square degrees, which also have a number of other advantages - relative cheapness, the ability to change the configuration, and (most importantly), since the dimensions of the detectors can be quite small, the use of high-temporal resolution instruments.  \cite{2011AN....332..714T,2010AdAst2010E..53B}. However, as can be seen from Figure~\ref{fig:exposure_limits} and Table~\ref{tab:surveys} of Appendix, only one multi-aperture system is currently being developed - Argus \cite{2022SPIE12182E..4HL}, which has a time resolution of 0.05 seconds. It consists of 38 (as a starting point, with at least 900 planned in the future) 20 cm telescopes with a total field of view of 344 (8000 in the final version) sq. degrees. Unfortunately, Argus is not able to change the configuration of the field of view, orienting all telescopes to one region of the sky, which will not allow it to be used as a single telescope with an effective diameter of about 5 meters. The remaining three instruments have the limiting dimensions of the fields of view (3-5 degrees) for a monolithic configuration in the Schmidt scheme and are equipped with sufficiently fast receivers. Nevertheless, it is clear that due to the small size of the fields of view when searching for fast transients, they can function only in the alert (follow-up) mode, like many large telescopes with standard fields of view of < 1 degree (see, for example, \cite{2022Univ....8..373T} and references therein). At the same time, this mode does not allow one to detect optical radiation synchronously with the fastest transients — gamma-ray bursts, fast radio bursts, and gravitational waves — for this, it is necessary to continuously monitor the celestial sphere using instruments that have the widest possible fields of view of hundreds and thousands of square degrees. In other words, it is necessary to independently detect and study fast optical transients and only afterwards look for their connection with events in other spectral ranges and of a different nature, comparing the time of occurrence and the position of both. We emphasize that the understanding of their origin, the choice of models that describe them, will largely be determined by the solution of this most complex problem of modern practical astrophysics. The areas of monitoring of gravitational waves, gamma and radio telescopes when searching for transients cover almost half of the area of the celestial sphere (the exception is the BAT detector of the Swift instrument with a field of view of about 5000 square degrees). This means that even with a relatively high accuracy (arc minutes - degrees) of determining the coordinates of an event for gamma detectors \cite{2022Univ....8..373T}, positioning optical telescopes according to an alert will take tens of seconds - minutes. At the same time, the duration of 90\% of gamma-ray bursts lies in the 20 ms - 100 s interval, and about 30\% of the events last less than 2 seconds \cite{1993ApJ...413L.101K}. The latter are the result of the merger of two neutron stars or a black hole and a neutron star and form a class of short gamma-ray bursts (SGRB), in contrast to long-duration GRBs (LGRB), caused by the collapse of some massive stars to form black holes \cite{2015PhR...561....1K, 2022Galax..10...38B}. Thus, in the alert mode, it is impossible to detect the optical companion of the short gamma-ray burst itself - gamma radiation is no longer detected by the time the telescope is pointed at the event localization region. When a long burst is detected, it is sometimes possible to begin optical observations of its location zone until the gamma-ray emission disappears. As a rule, in both cases, in the alert mode, it is possible to detect and study only the so-called afterglow – optical radiation resulting from the interaction of the plasma ejected during the generation of a gamma-ray burst with interstellar gas \cite{Meszaros1997}. 
 
It should be emphasized that so far, despite numerous works on observations and modeling of the phenomenon of gamma-ray bursts, no self-consistent theory is yet available. And it is generally accepted that the study of joint optical radiation with gamma emission (prompt emission), comparison of their characteristics and temporal structure can provide the key to solving this problem \cite{2015PhR...561....1K,2016A&AT...29..205C,2019A&A...628A..59O,2022hxga.book...31Y,2022ApJ...938..132P}. At the same time, out of 900 optically identified gamma-ray bursts, only 21 have scant optical information (only 1-3 brightness measurements) simultaneous with gamma-ray emission, and only six events have truly informative optical light curves obtained during the same period, usually in its final phase \cite{2019A&A...628A..59O}. Finally, for the single GRB 080319B (Naked Eye Burst) event, we detected a joint optical burst with a temporal resolution of 0.13 seconds during the entire GRB by the wide-field monitoring independent of gamma-ray telescope data \cite{2008GCN..7452....1K, 2008Natur.455..183R}. These unique observations with the wide- angle (600 sq. degrees) high-temporal resolution camera TORTORA \cite{2006NCimB.121.1525M} made it possible to establish the essential features of the mechanisms of gamma-ray burst generation \cite{2010ApJ...719L..10B}. The prototype of this camera is our similar instrument FAVOR (FAst Variability Optical Registrator), which was used for observations in 2003-2009 \cite{2005NCimC..28..747K, 2010AstBu..65..223B}.
 
Further, in the course of developing the program for wide-angle high temporal resolution monitoring, SAO RAS (together with KFU and OOO Parallax) created a 9-channel system with a field of view of 900 square degrees and a temporal resolution of 0.1 seconds, Mini- MegaTORTORA (MMT) \cite{2017AstBu..72...81B}. It was used to detect optical flares synchronous with the bursts GRB 160625B \cite{2018NatAs...2...69Z} (however, with a time resolution of 30 seconds) and GRB 210619B. In the latter case, radiation was recorded in 5 channels simultaneously with exposures of 1, 5, 10, and 30 seconds in white light (3 channels) and in filters B and V. The implementation of such a research mode is a fundamental feature of our approach, it allowed us to compare the structures of light curves in the gamma and optical ranges, build spectra in these ranges, and finally show that in this case the burst emission is due to a backward shock wave propagating in a relativistic jet \cite{2023NatAs...7..843O}.

Note that the MMT system is also equipped with polaroids with different orientations for measuring the linear polarization of optical flares, which should provide unique information about the physical properties of the bursts, namely, the structure of the emitting regions, the characteristics of the magnetic field, and the details of the mechanisms for generating radiation of different energies \cite{2015PhR...561....1K, 2020MNRAS.491.3343G, 2021Galax...9...82G}. Space-telescope measurements of the polarization of hard burst radiation have low accuracy (30 - 50\%) and temporal resolution \cite{2017NewAR..76....1M}, which does not allow one to choose between models, providing information only about the most general features of the phenomenon \cite{2016A&AT...29..205C}. Because of this circumstance, polarization studies of optical flashes accompanying gamma-ray emission are of particular importance. Such observations are practically absent - a small number of works on this topic contain the results of measurements of the polarization of optical afterglows, at the level of 10 - 20\%, but minutes - hours after the burst itself (\cite{2022A&A...666A.179B} and references therein).
In \cite {2015ApJ...813....1K} an upper limit for linear polarization at a level of 12\% have been obtained for the second episode of the GRB 140439A burst 2.5 minutes after the alert. Thus, the polarization of optical radiation simultaneous with the bursts, as well as its spectral characteristics, have yet to be investigated in high-resolution wide-angle surveys.

At present, one of the unsolved (and most important) problems of observational astrophysics is the detection of optical companions of fast radio bursts - the choice of adequate models of these phenomena from a large number of the ones considered requires the study of their possible optical counterparts, which, unlike gamma-ray bursts, are the subject of only theoretical constructions. Some of the several dozen models predict the generation of sufficiently bright optical flares \cite{2017MNRAS.468.2726K,2018NatAs...2..845B,2019A&ARv..27....4P}. On the other hand, the frequency of occurrence of fast radio bursts throughout the sky can reach several thousand per day \cite{2013Sci...341...53T}, and, consequently, the optical flares accompanying them can be quite frequent.  At the same time, as we know, unlike radio observations, optical surveys to search for these events with a high (1 ms) time resolution are not carried out at the moment. As a rule, studies are focused on searching for stationary optical burst hosts \cite{2018PrPNP.103....1K}, or searching for millisecond optical flashes in repetitive bursts. About 20 such objects, which erupted about 200 times, were found among the almost 800 bursts proper. Their coordinates were determined, which made it possible to observe the regions of localization of these sources using large telescopes (see, for example \cite{2018MNRAS.481.2479M} and \cite{2017MNRAS.472.2800H}), in which optical flares were not detected - the same result was obtained by us with a 6-meter telescope \cite{2022Photo...9..950V}. Gravitational wave events can also be accompanied by optical flares ranging from a few milliseconds to a few seconds, especially when they come from merging neutron stars, like the recently discovered event GW 170817 \cite{2018ApJ...867...18N, 2018ApJ...852L...5M}. At the same time, since the accuracy of localization of gravitational waves is hundreds of degrees, such flares can only be detected by monitoring the celestial sphere independent of the data of gravitational detectors using instruments with comparable fields of view and millisecond time resolution.  
Moreover, in the constantly updated catalog of fast radio burst models \cite{2016PASA...33...45P}, probable deep connections between them, the gravitational waves and gamma-ray bursts are clearly visible, based on the universal cause of these phenomena - the interaction and evolution of relativistic objects.  It is all the more important to look for signatures of similar phenomena in the optical range using wide-angle monitoring of high temporal resolution. Based on the above considerations, and on the experience of developing and using the FAVOR, TORTORA, Mini- MegaTORTORA tools, we proposed the concept of a multi-telescope complex with high temporal resolution for wide-angle monitoring of the celestial sphere - SAINT (Small Aperture Imaging Network Telescope) \cite{2013PhyU...56..836B, 2017ASPC..510..530B, 2022Photo...9..950V, 2022AstBu..77..495V} using telescopes with a diameter of 30-50 centimeters, with the limiting magnitude at the level of single- component instruments (see Fig.~\ref{fig:exposure_limits} and Table~\ref{tab:surveys}) and capable of detecting and investigating non-stationary objects at times from milliseconds to months-years. This paper is devoted to the development of this project using modern devices, the choice of the design of the complex, its components, mode of operation, and the assessment of its parameters.

\section{Small Aperture Imaging Network Telescope (SAINT) project}
\label{sec:saint}

The performed analysis demonstrates the obvious need to create a multi-channel optical monitoring system that would be capable of conducting complex searches and studies of non-stationary objects and phenomena within the framework of universal instrumental and methodological approaches and would not have the disadvantages of existing systems.

The features of this kind of instrument should be as follows:

\begin{itemize}
    \item a large field of view of several hundred degrees, which requires a multi-channel design;
    \item high temporal resolution in the range of tenths to hundredths of a second;
    \item detection limit of 18 -- 20 mag  at time scales of 20-30 seconds to be at the level of the limits of existing survey telescopes (Fig.~\ref{fig:exposure_limits});
    \item a combination of monitoring (wide-field) and follow-up (narrow-field) modes with their rapid change;
    \item the possibility of obtaining maximum information (spectral, polarimetric, photometric) about non-stationary objects in follow-up mode;
    \item processing of accumulated information in real time to detect, characterize and classify transient phenomena, and make decisions on the transition to the follow-up mode;
    \item preservation of all raw and reduced data, as well as maintenance of databases obtained as a result of its \textit{a posteriori} analysis;
    \item complete robotization of the complex operation using information from external sources (meteo station, network, other instruments) and a system to control its condition.
\end{itemize}

\begin{figure}[t]
\centering
\centerline{
    \resizebox*{1\columnwidth}{!}{\includegraphics{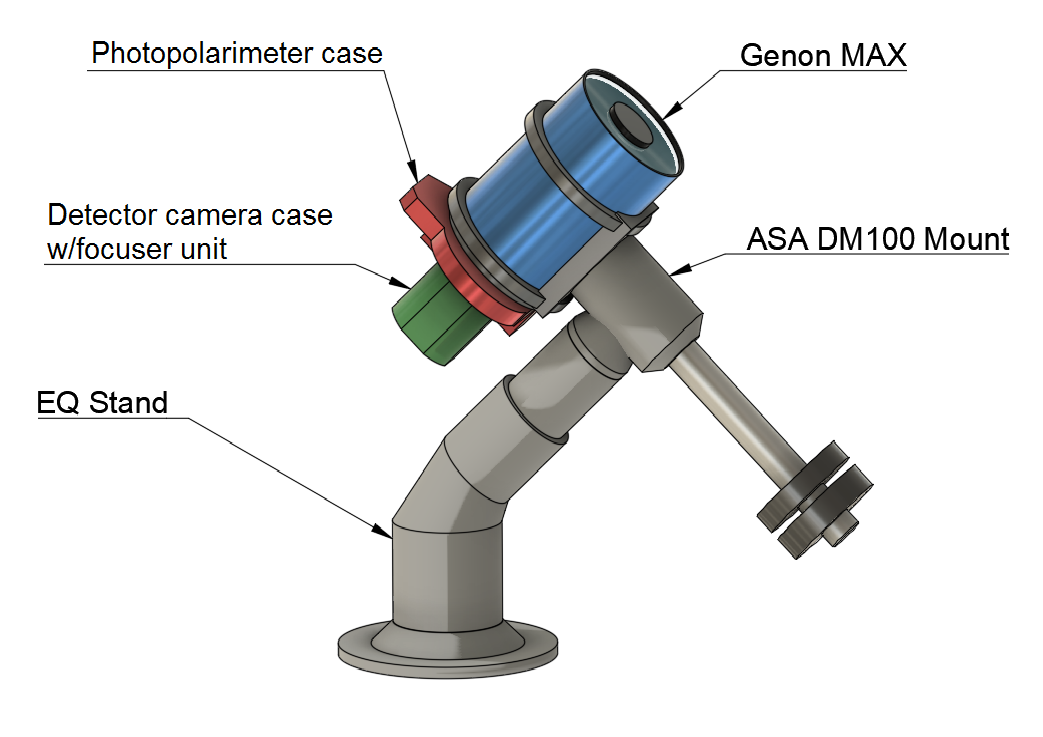}}
}
\caption{Schematic view of a single assembled channel of SAINT complex.
\label{fig:channel}}
\end{figure}

As a practical implementation of these principles, we propose the project of 
SAINT (Small Aperture Imaging Network Telescope) -- 12-telescope complex with high temporal resolution, built mostly from off-the-shelf commercially available components that may be implemented in about \$2M cost (plus additionally about \$500K for labor costs). 
Its overall parameters are listed in Table~\ref{tab:saint}, schematic view of a single channel is shown in Figure~\ref{fig:channel}, and details on individual components and details of its operation modes are given below.

\begin{table}
\begin{center}
\caption{Main characteristics of the SAINT complex.\label{tab:saint}}
\begin{tabular}{lc}\hline
Number of telescopes / mounts & 12 / 12 \\
Telescope aperture & 300 mm \\
Telescope focal length & 450 mm \\
Optical scheme & Schenker-Terebizh \\
sCMOS array size & 49.5 x 49.2 mm \\
Fullframe readout speed & 54 fps\\
Pixel scale & 5.5 $''$/pix \\
Telescope FOV & 6.3$^{\circ}$ x 6.3$^{\circ}$ = 39 sq.deg.\\
System FOV & 468 sq.deg. \\
\hline
\end{tabular}
\end{center}
\end{table}

\subsection{Telescope}
\label{sec:telescope}

The basic component of the complex is GENON Max -- a catadioptric two-mirror Schenker-Terebizh telescope with five corrective lenses \cite{2011AN....332..714T} with a diameter of 30 cm (F/1.5) and a field of view of about 40 square degrees (Table~\ref{tab:genon}).
In its Cassegrain focus
a multimode photopolarimeter of a rather complex design is installed, capable of operating both in wide-field monitoring and in the study of a single object (Section~\ref{sec:photopolarimeter}). 

\begin{table} 
\begin{center}
\caption{Main characteristics of the GENON Max telescope (design by G. Borisov).}
\label{tab:genon}
\begin{tabular}{lc}
\hline
Scheme with Cassegrain focus & Schenker-Terebizh \\
Diameter, mm & 300 \\
Focal length, mm & 450 \\
Wavelength range, $\AA$ & 4500 - 8000 \\
The coefficient of central screening by diameter & 0.50 \\
Linear diameter of the field of view, mm & 70 \\
Angular diameter of the field of view, deg. & 8.9 \\
Relative illumination across the field, center - edge of the field & 1.0 - 0.7 \\
Telescope weight (tube with optics assembled), kg & 29 \\
Optical system length, mm & 430 \\
\hline
\end{tabular}
\end{center}
\end{table}

\subsection{Mount}
\label{sec:mount}

The telescopes are mounted on ASA DDM100 mounts based on direct drive technology (see Table~\ref{tab:asa}).
This design has a number of important advantages  listed below, which are especially significant for the proposed complex.

\begin{itemize}
    \item High acceleration and, as a result, high pointing speed up to 50$^{\circ}$/s, which will allow changing the field of view of the complex and the observation mode in several seconds. 
    \item Due to the absence of drive belts and gears, the rotation of the axes is very uniform and silent, which ensures the accuracy of guidance and the absence of vibrations, which determine the high stability of the position of sources in the focal plane of telescopes.
    \item Due to the absence of gearboxes, the absence of mechanical backlash and hysteresis is ensured. For the same reason, moving parts wear out little, which increases the period of maintaining the stability of the mount’s characteristics.
    \item All movable structural elements are equipped with 28-bit encoders and feedback systems, which ensures the accuracy of rotation angles and their determination up to 0''.004.
    \item Maintaining the stability of the functioning of all elements of the mount and its control is provided along with feedback systems by a high-speed controller on line with a computer.
\end{itemize}

\begin{table}
\begin{center}
\caption{ASA DDM100 mount specifications.}
\begin{tabular}{lc}
\hline
Type & Equatorial \\
Motor & Direct drive \\
Max load & 100 kg \\
Weight (without counterweight) & 56 kg \\
Power supply & 24 Volts \\
MAX slew speed & 50 deg/s \\
Pointing Accuracy (20° to 85°)  & $<8''$ rms \\
Tracking Accuracy (20° to 85°)  & $<0.25''$ rms \\
\hline
\end{tabular}
\label{tab:asa}
\end{center}
\end{table}

\subsection{Detector}
\label{sec:detector}

Every channel of SAINT is equipped with a photopolarimeter (Section~\ref{sec:photopolarimeter}), which is the primary instrument in the SAINT complex. The detector used in it is Andor sCMOS Balor 17F-12 camera, whose characteristics are listed in Table~\ref{tab:balor}.

\begin{table}
\begin{center}
\caption{Characteristics\textsuperscript{*} of the Andor Balor sCMOS camera.}
\label{tab:balor}
\begin{tabular}{lc}
\hline
Sensor size W x H, mm (px) & 49.5 x 49.2 ( 4128 x 4104 ) \\
Cell size, $\mu$m & 12 x 12 \\
Readout noise, e- & 2.9 \\
Full well depth, e- & 80000 \\
Frame rate of full frame reading, Hz & 54 \\
Maximum quantum efficiency (at 6000 $\AA$), \% & 61 \\
ADC, bit & 16 \\
The highest frame rate for the reading area 512x512, Hz & 431 \\
\hline
\end{tabular}
\\
\footnotesize \textsuperscript{*}All characteristics are given according to official technical specifications from Andor website, available at
\url{https://andor.oxinst.jp/assets/uploads/products/andor/documents/andor-balor-17F-12-specifications.pdf}.
\end{center}
\end{table}

This detector currently has the best combination of a large size of 16.9 Mpix (70 mm diagonal), high temporal resolution when reading a full frame (about 20 ms) and a fairly low readout noise (2.9 e-), which is unattainable in CCD matrices for several data transmission channels \cite{2020INASR...5..236S}. A sufficiently large pixel size (12 µm) allows using the maximum temporal resolution in observations even with relatively low image quality ($>$ 1$''$), nevertheless concentrated in one pixel, which keeps the readout noise to a minimum. At the same time, a high degree of signal response linearity ($>$99.7\%) is combined with a large electron well depth (80000) and a special way of amplifying and digitizing the signal, which ensures its continuous recording in the maximum dynamic range. These features make it possible to detect sources with an intensity from the read-out noise level to the saturation limit without distortion in a single camera frame, which is fundamental in wide-field observations.
Moreover, reducing the read-out area (Region of Interest, ROI) (Table~\ref{tab:balor_fps}) allows to improve the temporal resolution down to 2.5 or 0.6 ms which may be used in the regime of targeted observations of individual objects.


\begin{table}
\begin{center}
\caption{Maximum frame rate depending on the size of a read-out region of a sensor (region of interest, ROI).}
\begin{tabular}{cc}
\hline
\textbf{ROI size, pixels}	& \textbf{Max frame rate, fps}\\
\hline
4128 x 4104 & 54 \\
2048 x 2048 & 108 \\
1920 x 1080 & 205 \\
1024 x 1024 & 216 \\
512 x 512 & 431 \\
128 x 128 & 1684 \\
\hline
\end{tabular}
\label{tab:balor_fps}
\end{center}
\end{table}

The microlens array on top of the sensor in Balor camera allows lossless operation for the relative apertures of up to F/0.3 at cone angle up to  110$^{\circ}$, which makes it possible to avoid light losses even at the outer margins of the telescope field of view at aperture ratio of F/1.5. Important point is what the camera allows to timestamp and synchronize the acquired frames using the signal from GPS receiver which is crucial for combination of data from different channels pointed towards a single target. On the other hand, in contrast to traditional CCD sensors, the question of long-term spatio-temporal stability of individual pixels of CMOSes pose potentially significant problem for the co-addition of images from such detectors, as it introduces additional per-pixel noise that cannot be mitigated by classical calibration methods (bias and dark frame subtraction, non-linearity correction and flat fielding). The limits imposed by these low-frequency effects on frame co-addition are still poorly studied, and have to be thoroughly investigated prior the decision on exact strategy of frame co-addition in the operation of the complex.

\subsection{Photopolarimeter and different modes of operation}
\label{sec:photopolarimeter}

The primary instrument of every SAINT channel, and the only custom-built part of it, is a photopolarimeter which provides observations in both monitoring and research modes. 
It is attached to the Cassegain focus of the telescope and consists of two parts -- an optical-mechanical unit and a camera unit, its general view with installation on the telescope is shown in Figure~\ref{fig:channel}.

The camera unit houses the detector, which is mounted on sliding rails that allow it to move along the optical axis, thereby focusing images on the sensor. The hoses of the water cooling system of the chamber, not shown in the figure, are inserted into the rear end of the casing of the unit.

\begin{figure}
\centering
\centerline{
    \resizebox*{1\columnwidth}{!}{\includegraphics{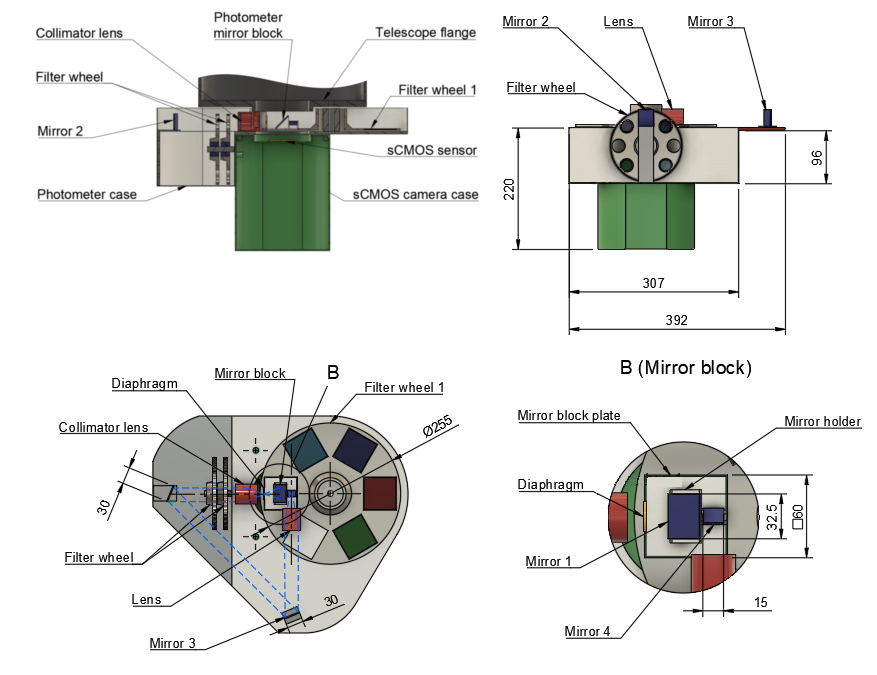}}
}
\caption{Schematic view of a photopolarimeter installed in the Cassegrain focus of every channel of SAINT complex.
\label{fig:photopolarimeter}}
\end{figure}

\begin{figure}
\centering
\centerline{
    \resizebox*{1\columnwidth}{!}{\includegraphics{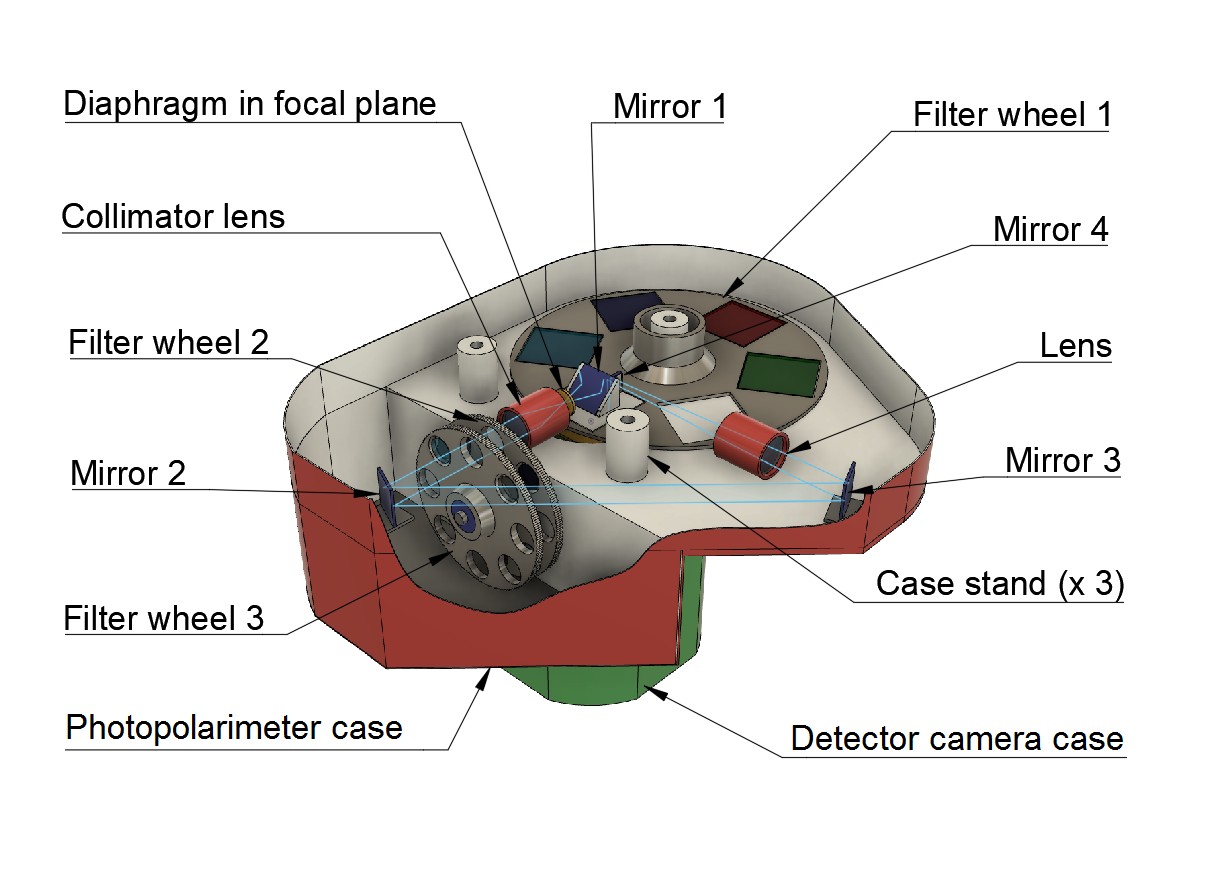}}
}
\caption{Internal view of the photopolarimeter and the course of rays in the follow-up mode.
\label{fig:photopolarimeter2}}
\end{figure}

\begin{figure}
\centering
\centerline{
    \resizebox*{1\columnwidth}{!}{\includegraphics{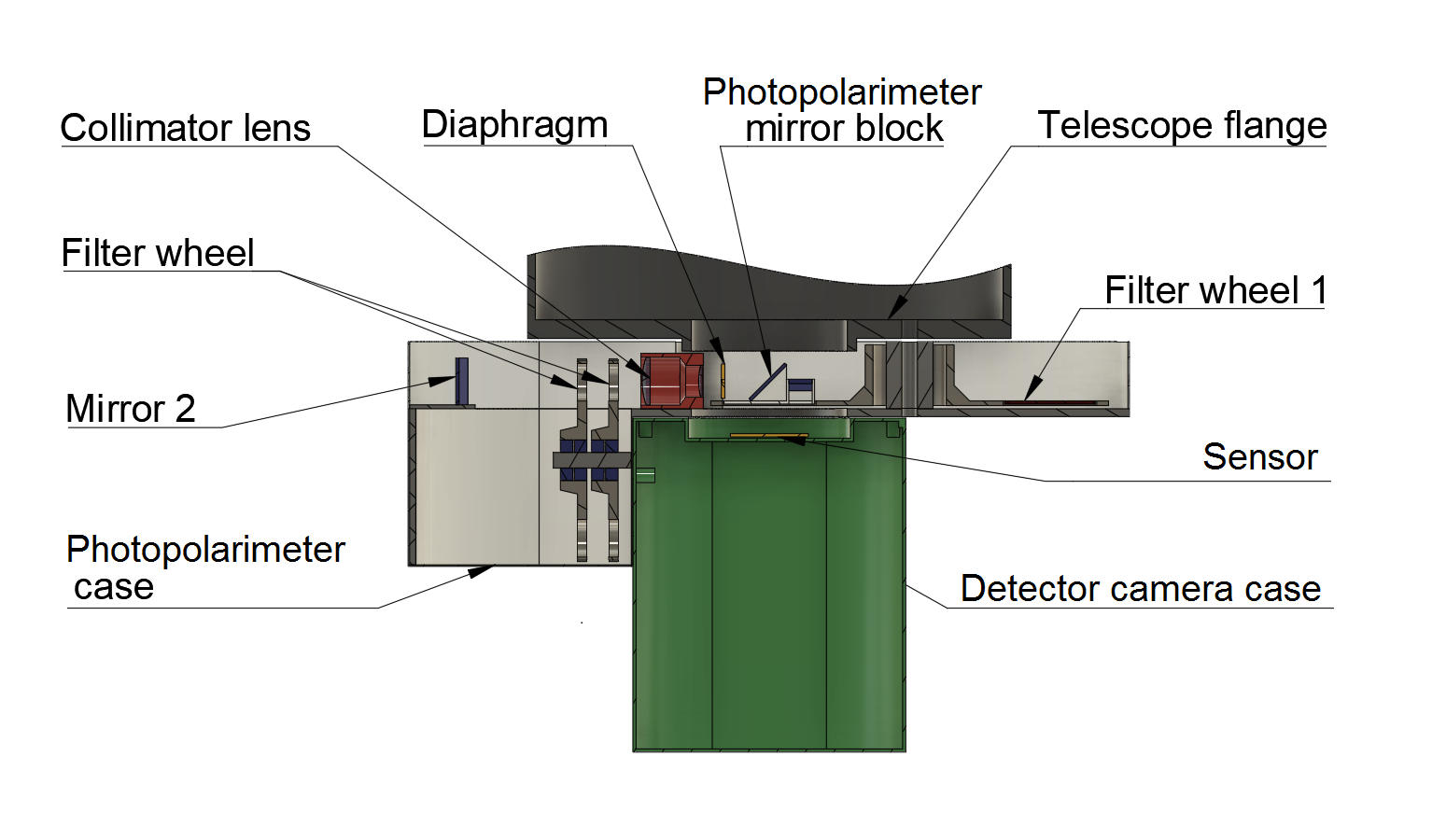}}
}
\caption{Longitudinal view of the SAINT photopolarimeter.
\label{fig:photopolarimeter3}}
\end{figure}

The structure of the photopolarimeter is shown in the diagrams of Figure~\ref{fig:photopolarimeter}, its internal view is in Figure~\ref{fig:photopolarimeter2}, its longitudinal view is in Figure~\ref{fig:photopolarimeter3}. Control electronics units, power supply, motors that ensure the movement of moving elements of the device, and connecting cables are not shown here.

The front part of the photopolarimeter, located in the gap between the back focus of the telescope and the entrance of the camera, contains the following components: a turret with $griz$ filters, an empty window, and a narrow-field block containing mirrors 1 and 4, as well as a field-of-view diaphragm (the optical axis of the telescope passes through the centers of the turret holes); input (COMPUTAR M2514-MP2 with a focal length of 25 mm and aperture ratio of 1:1.4) and output (COMPUTAR V5014-MP with the focal length of 50mm) lenses of the  collimator; mirror 3.

The side part of the photopolarimeter is located outside the camera casing. It contains: a turret with narrow-field $griz$ filters and an empty window, as well as a turret coaxial with it with three polaroids of different orientations (with polarization plane rotated by 120$^{\circ}$ in respect to each other) and an empty window; mirror 2.

During observations in the wide-field (monitoring, or primary survey) mode, an empty turret window 1 or one of its filters is installed on the optical axis of all 12 telescopes (depending on the specific tasks of the survey, observations of different areas with different filters, etc. are possible). The cameras acquire a sequence of full-frame images (4128 x 4104 pixels, or 6.3$^{\circ}$x6.3$^{\circ}$ each) with a rate of 54 fps, and the entire complex monitors the overall field of view at 470 sq. degrees.

In the narrow-field (research) mode, a block with mirrors 1, 4 and a field diaphragm measuring 11$'$ x 11$'$ (1.45 x 1.45 mm) of turret 1 (see Fig.~\ref{fig:photopolarimeter}) is installed on the optical axis, while mirror 1 focuses the redirected beam on the diaphragm placed at the focus of the input lens of the collimator. The parallel beam constructed by it passes through one of the $griz$ filters (or empty window) of turret 2 and one of the polaroids (empty window) of turret 3. After being reflected by mirrors 2 and 3, the parallel beam is focused by the output lens of the collimator and redirected by mirror 4 to the sensor of the camera. As a result, the light from the 11$'$ x 11$'$ sky region with a linear size of 3 x 3 mm (240 x 240 pixels) is imaged with a scale of 2.75$''$/pix, twice the original (5.5$''$/pix) --  the scale is determined by the focal length  of the collimator output lens.

\begin{figure}
\centering
\centerline{
    \resizebox*{1\columnwidth}{!}{\includegraphics{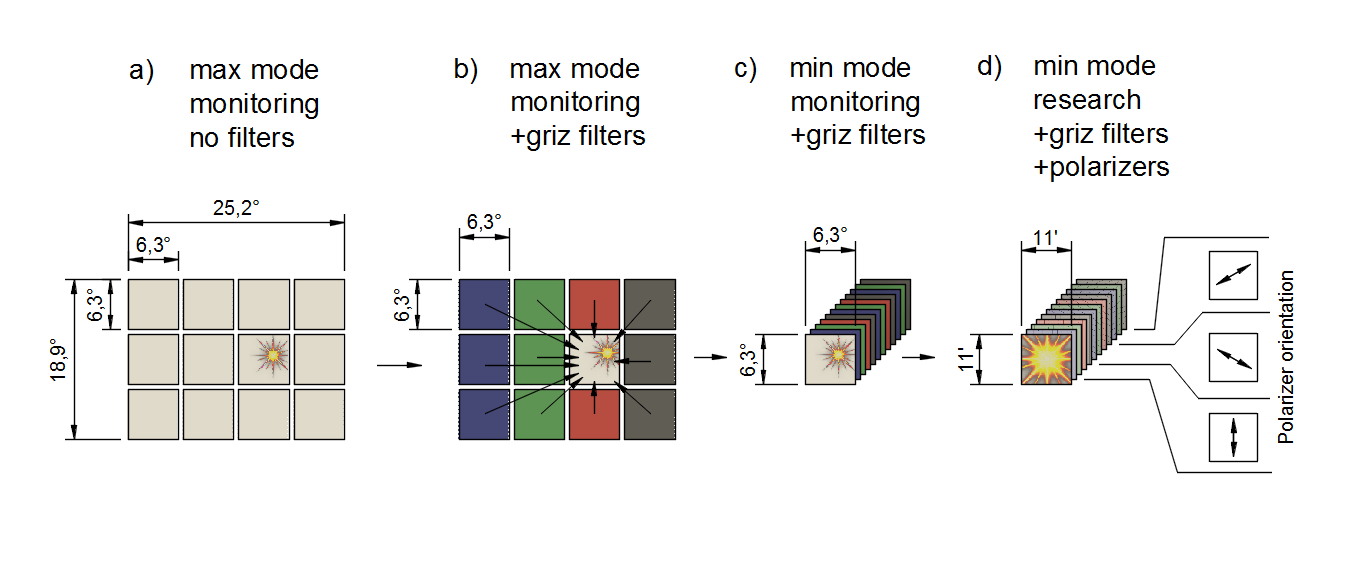}}
}
\caption{Change of observation modes of the complex after detecting an optical transient.
\label{fig:modes}}
\end{figure}

This image can be recorded with a time resolution of 20 to 2.5 ms, and its central 6$'$ x 6$'$ part (128 x 128 pixels) -- with a resolution of 0.6 ms (see Table~\ref{tab:balor_fps}). The transition from monitoring to research mode takes a few seconds, and is determined by both the rotation speed of the photopolarimeter turrets, and the repointing speed of the mount.
In particular, when an optical transient is detected during real-time monitoring, the following operations are performed:

\begin{itemize}
    \item its characteristics are determined (initial brightness, characteristic duration, structure of the light curve);
    \item rough initial classification is performed (as one of noise, meteor, satellite, new or already known astrophysical object classes);
    \item the follow-up mode for every channel is selected (photometry and/or polarimetry, filters to use, exposure time);
    \item in parallel, all telescopes are repointed to the source, placing it in a small central field (with the exception of the instrument that detected the transient, it retains the original mode used for detection, in order to keep uninterrupted sequence of data);
    \item observations start.
\end{itemize}
Figure~\ref{fig:modes} illustrates this process.

Table~\ref{tab:limits} shows the estimated detection limits of the complex in different modes and with different time resolutions. Estimates were made for a single exposures in white light at a characteristic wavelength of 5500 $\AA$, a sky background of 21.5 mag/sq.arcsec, an optical and atmospheric transmission of 0.25, the characteristics of the telescope and detector from Tables~\ref{tab:genon} and \ref{tab:balor}, and assuming that the object flux is fully contained inside single pixel (i.e. significantly undersampled PSF).
It should be noted that the real limits when using filters and polaroids will be 0.5 -- 1.0 magnitude worse.

\begin{table}
\caption{Detection limits in different modes.\label{tab:limits}}
\renewcommand{\arraystretch}{1.2}%
\begin{tabular}{rcccccccccc}
\hline
\textbf{Mode} & \textbf{Scale} & \textbf{Size} & \textbf{Mode$^{*}$} & \textbf{FOV} & \multicolumn{6}{c}{\textbf{Detection limit at exposure}} \\
& \textbf{FOV} & pixels & & sq.deg. & 0.6ms & 2.5ms & 20ms & 1s & 30s & 20 min \\
\hline

\multirow[c]{2}{*}{Monitoring} & 5.5$''$/pix & \multirow[c]{2}{*}{4128 x 4104} & Max & 468 & & & 12.6 & 16.5 & 18.9 & 20.9 \\


& 6.3$^{\circ}$x6.3$^{\circ}$ &  & Min & 39 & & & 14.0 & 17.9 & 20.2 & 22.2 \\

\hline

\multirow[c]{2}{*}{Follow-up} & 2.75$''$/pix & \multirow[c]{2}{*}{240 x 240} & Max & 432 & 8.8 & 10.3 & 12.6 & 16.7 & 19.5 & 21.6 \\


& 11$'$x11$'$ &  & Min & 36 & 10.2 & 11.6 & 14.0 & 18.0 & 20.8 & 23.0 \\

\hline
\end{tabular}
\noindent{\footnotesize{
* Possible modes of operation: \\
- Max -- telescopes are pointed at different areas (12 x 39 sq. deg.) \\
- Min -- telescopes are pointed at one area (39 sq. deg.)
}}
\end{table}

\subsection{Shelter and infrastructure of the complex}
\label{sec:shelter}

\begin{figure}[t]
\centering
\centerline{
    \resizebox*{1\columnwidth}{!}{\includegraphics{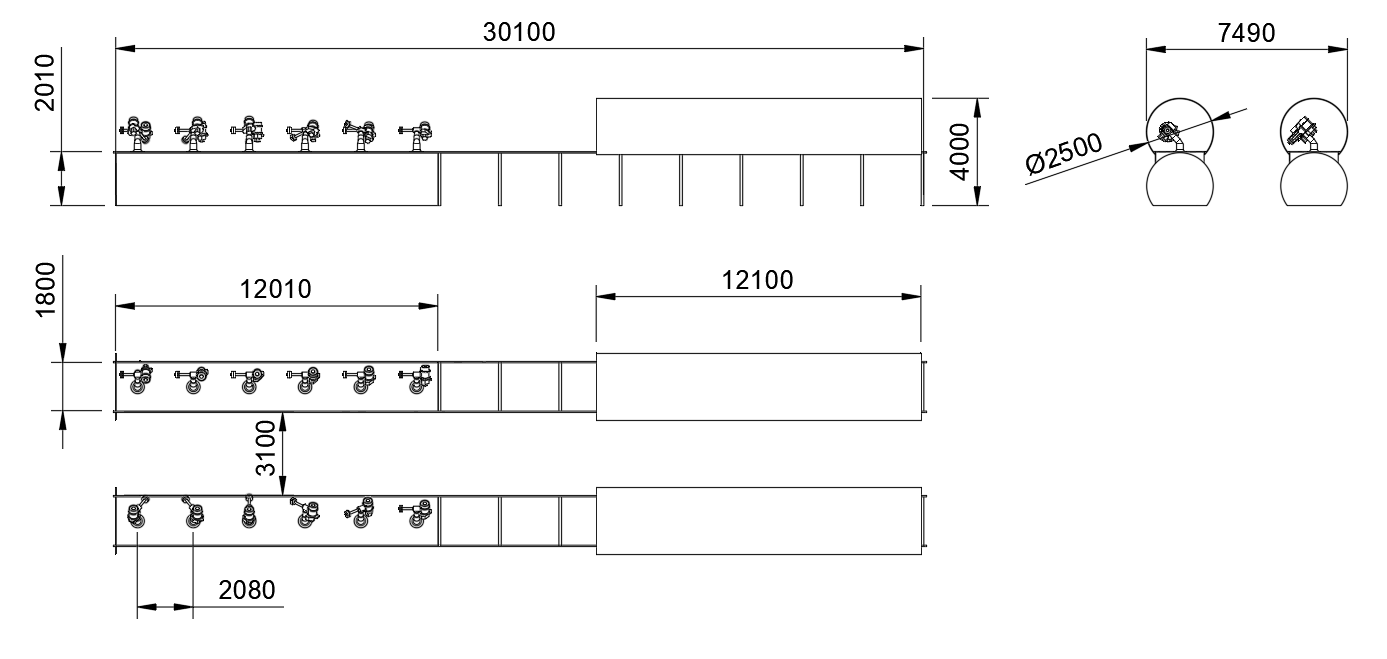}}
}
\centerline{
    \resizebox*{1\columnwidth}{!}{\includegraphics{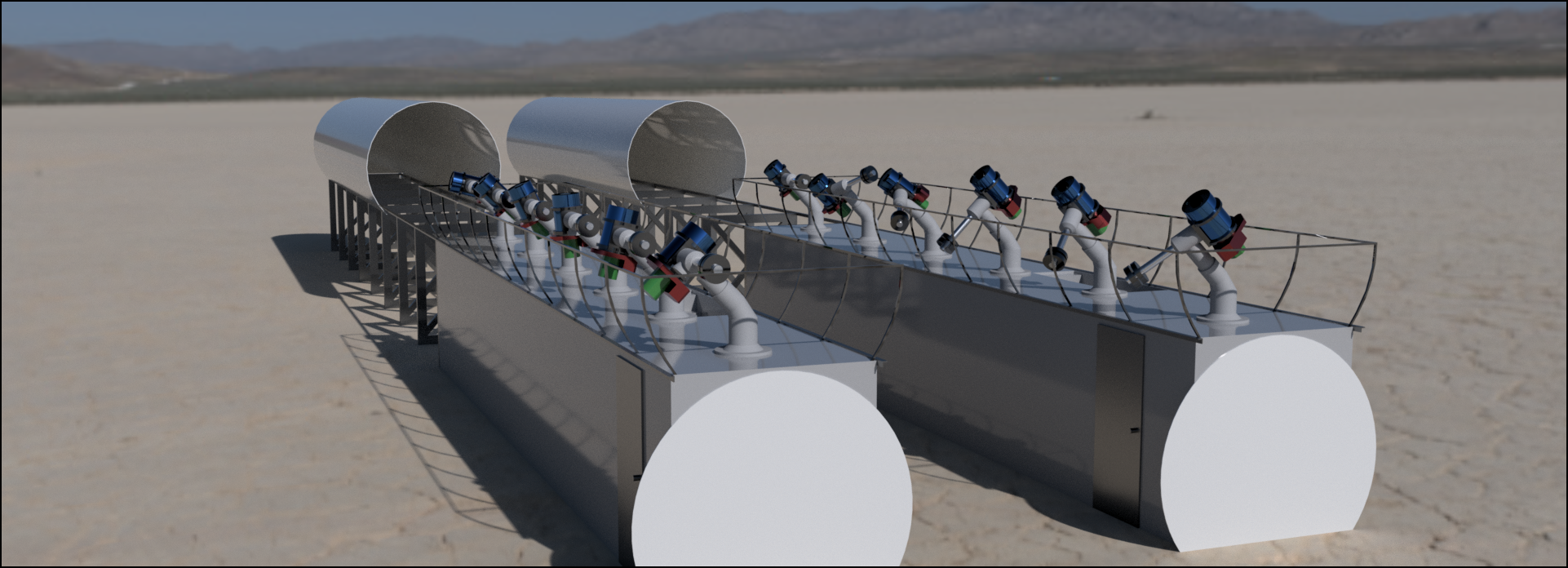}}
}
\caption{The scheme and general view of the shelter of the SAINT complex.
\label{fig:shelter}}
\end{figure}  

Twelve telescopes of the SAINT complex are installed in two cylindrical shelters housing six channels each (see Fig.~\ref{fig:shelter}). They consist of a reinforced concrete supporting slab mount and a steel truss frame sheathed with sandwich panels on which it is mounted. Its free part serves as a flyover to accommodate the cylindrical roof of the shelter in the working state of the complex, which, moving along rail guides, closes the plate with the telescopes in the non-working state. The frame is fixed on a reinforced concrete base covered with metal-plastic cladding, installed on a concrete pad dug into the ground. 

The internal volumes of the frame are the technical compartments of the SAINT complex.
They house the following equipment and infrastructure components:
\begin{itemize}
    \item power supply input unit of the complex network, including power drives of all its mechanical elements, electronic components of the telescopes and instruments, computer cluster, uninterruptible power supplies;
    \item system for providing water cooling of detectors;
    \item climate control system based on dehumidifiers COTES CR300.
\end{itemize}

Both parts of the shelter are equipped with video cameras for monitoring the internal and external space. The complex is also equipped with a meteorological system, a fish-eye all-sky camera and an IR cloud sensor, which allows to automatically monitor weather conditions and make a decision on the start and end of observations.

The movable roofs of the shelters are covered with a sun-protection coating, and flat screens with uniform illumination are placed on their inner surface to calibrate the sensitivity of the detectors.

\section{Data acquisition, processing and storage}
\label{sec:dataprocessing}

During the primary monitoring mode of operation, the complex will acquire the data for every sky field (470 square degrees) for 20 minutes in a row, thus covering up to 12700 square degrees in a typical 9-hour observational night. The data will be processed in real time, and a number of data products (transients on various time scales, co-added images with different temporal resolution, etc) will be derived and stored.

As both the computational hardware and data analysis algorithms evolve very quickly, it is not possible (and not necessary) to specify the IT infrastructure of SAINT complex here in full details as we did for the telescope itself. Therefore, below we will outline its generic structure, requirements and possible approaches for handling an enormous amount of data the complex will produce.


In general, the operation of the complex will be carried out using the software that is based on that we used in observations with the FAVOR, TORTORA and Mini-MegaTORTORA instruments (see \cite{2017AstBu..72...81B} and references therein). On the shortest time scales, where timing constraints are the most demanding, transient detection will be performed in real time using fast image subtraction methods that are based on sufficient local stability of image PSF, as described e.g. in \cite{2010AdAst2010E..40K}. On longer time scales, where PSF stability cannot be ensured, the methods developed for image subtraction with PSF matching developed over last 20 years will be used, such as \cite{1998ApJ...503..325A} family of methods, ZOGY \cite{2016ApJ...830...27Z} algorithm, or SFFT \cite{2022ApJ...936..157H}. These methods generally allow parallelization using modern GPU hardware, and thus may be expected to run fast enough to be used for analyzing the data in real time on minutes to hours time scale. An alternative to image subtraction methods may also be implemented using convolutional neural networks (CNN, see e.g. \cite{2018MNRAS.476.5365S} or \cite{2023AJ....166..115A}).
CNNs will also be employed in order to reliably separate the image subtraction artefacts, cosmic rays, as well as e.g. effects of stellar scintillations on short time scales from \textit{bona fide} transients using the methods developed in e.g. \cite{2017ApJ...836...97C, 2017MNRAS.472.3101G, 2019PASP..131j8006C, 2022A&A...664A..81M}

Image co-addition will also allow simultaneously achieving highest possible temporal resolution (on individual frames) and going for deeper objects (in running coadds of various length, corresponding to e.g. every 100 or 1000 original images). Unfortunately, without thorough laboratory study of the stability of the detector it is impossible to define the limits for such co-addition, as at some point the effects of pixel parameter drifts specific for CMOSes  will be more significant than the further sensitivity gain. However, we may foresee that implementing some tailored observational strategy in the data sequence -- e.g. dithering using some pre-defined pattern -- will allow to mitigate such effects to some degree.

As a minimal architecture for the data processing cluster for SAINT we propose a hierarchical scheme with two identical computers per telescope channel -- one for data acquisition from the CMOS and real-time image processing, second for the channel hardware and high-level controls, intermediate data storage and image processing -- plus a central computer that handles the overall operation of the complex, including survey scheduling and reaction to the detected transients, communication with external networks, as well as various  databases. Such architecture also allows to use all cluster machines for a more computationally-intensive data processing tasks during day time, when no observations are performed and no real-time data processing is necessary.

Let's outline the basic requirements for data processing and storage infrastructure. They are defined by the extremely large amount of data being acquired from several large format detectors (16.9 millions of pixels each for Balor camera) routinely operating at the frame rates up to 54 frames per second (or even up to 1684 fps in follow-up regime), thus amounting to 1.8 gigabytes of data per second for every channel, or up to 800 Terabytes per night for the whole complex. As a minimal realistic configuration using present-day commercially available hardware, we may use a camera connected to the PC through a four-channel CoaXPress interface, and a RAID-0 array consisting of five SATA-III harddrives 20 terabytes each. Due to parallel writing, the speed of up to 4 gigabytes per second may be achieved for such RAID configuration, which is sufficient for handling the real-time data flow. Longer term data storage will require datacenter-class hardware which is also commercially available.

\section{Expected results}
\label{sec:results}

\begin{figure}[t]
\centering
\centerline{
    \resizebox*{1\columnwidth}{!}{\includegraphics{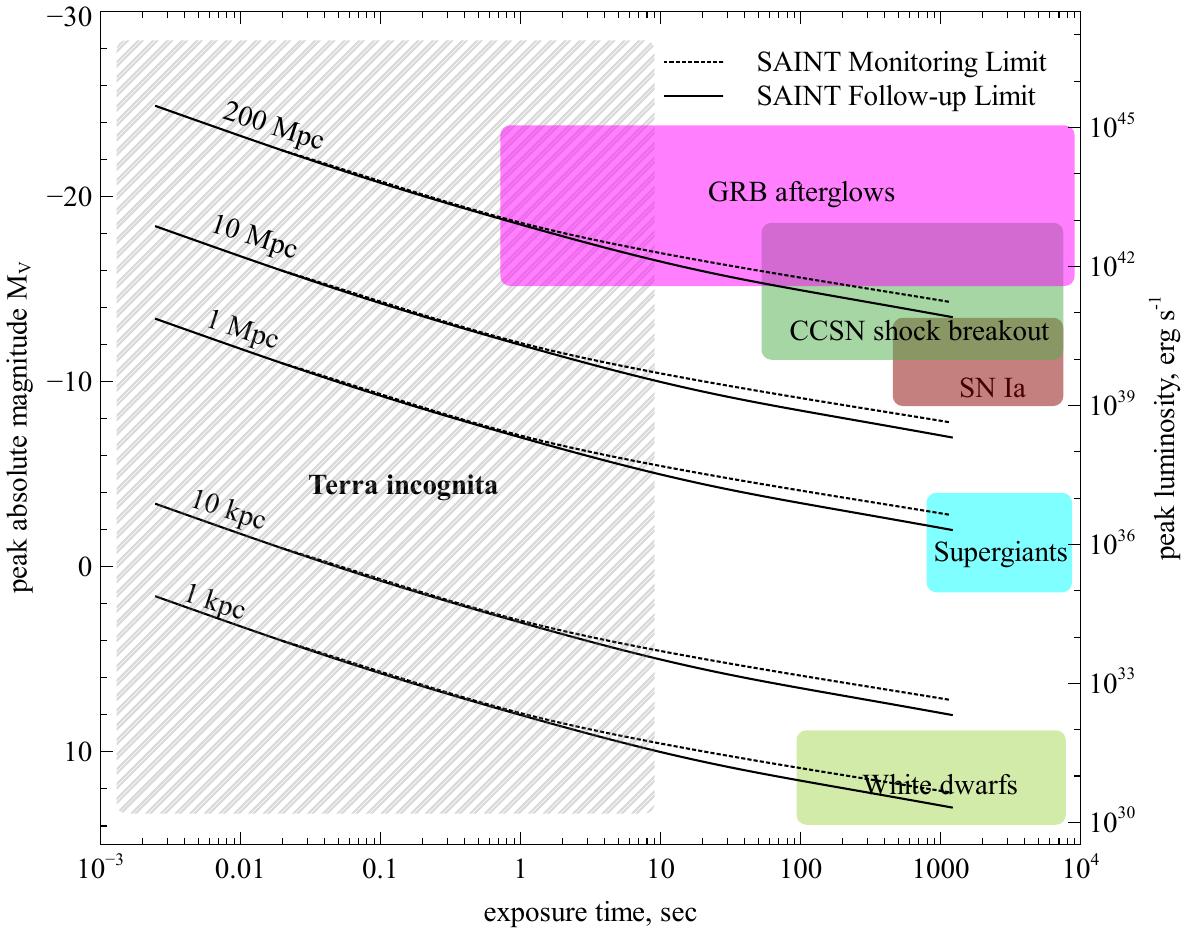}}
}
\caption{Detection limits of the SAINT according to the Table~\ref{tab:limits} for various distances, timescales and peak luminosities.
\label{fig:saint_limits}}
\end{figure}

The proposed complex is able to operate in still poorly studied region of the parameter space shown in Figure~\ref{fig:exposure_limits}, and thus may provide a number of results related to previously unknown classes of rapid optical transients, as well as a vast amount of data on known ones. Below we will briefly outline some of the fields where such instrument may provide important developments.

Flaring stars are now being routinely discovered in modern time domain sky surveys like Kepler \cite{2016ApJ...829...23D}, TESS \cite{2020AJ....159...60G, 2023A&A...669A..15Y}, EvryScope \cite{2019ApJ...881....9H}, NGTS \cite{2023MNRAS.525.1588J}, ZTF \cite{2021MNRAS.505.1254K} or Tomo-e~Gozen \cite{2022PASJ...74.1069A}  what perform high cadence continuous observations of the fixed sky regions. However, most of these experiments do not provide temporal resolution better than half a minute, or in rare cases -- better than a second, and are being performed in a single photometric filter that somehow reduces the scientific content of collected statistical results. Thus, using specifically tailored monitoring mode with SAINT what is possible due to its multi-channel architecture -- i.e. simultaneous multicolor observations -- would allow to augment the statistics of white-light flares with at least some estimations of their temperature. Moreover, employing the polarimetric mode for targeted (or follow-up, for the flares detected by real-time processing pipeline) observations would also allow to place statistically reliable upper limits on (or even get a detection of) the polarized components of stellar flares of various durations, including the shorter ones not typically resolved by existing surveys (see also Figure~\ref{fig:saint_limits} for details).

\cite{2022PASJ...74.1069A} reported 22 flares detected in 40 hours of 1 second cadence observations with Tomo-e~Gozen camera (21 square degree simultaneous field of view).
While the depth of SAINT survey is not as good, due to superior field of view we may expect up to 12 flares per hour in a wide-field mode, thus getting significantly larger amount of flashes, covered with better temporal resolution.

The observations of faint meteors with the Mini-MegaTORTORA system \cite{2017AstBu..72...81B} typically resulted in detection of several hundreds events per observational night, with maximum brightness down to 8-10 magnitdues, which is 2--4
magnitudes deeper than the limits of most modern meteor detection experiments (see, for example, \cite{2013Icar..225..614W, 2017PSS..143..116}) and
comparable to the capabilities of the Tomo-e~Gozen camera \cite{2018SPIE10702E..0JS}. Such faint, and even fainter meteor events are caused by micrometeoroids -- particles
of interplanetary (interstellar?) dust with masses of 0.1$\mu$g--0.1g \cite{1998SSRv...84..327C} -- that, according to direct transatmospheric
studies, make up the majority of the cosmic matter falling to the Earth, 100--200
tons/year \cite{1993Sci...262..550L}. On the other hand, observations of meteors give an order of magnitude smaller volume \cite{2012ChSRv..41.6507P} -- this discrepancy is most probably due to an underestimation of the
role of ablation and fragmentation processes of micrometeoroids, as well as inaccuracy in
determining their velocities \cite{2012ChSRv..41.6507P}. These problems
are largely associated with the difficulty of interpreting radar observation data, which is
the main means of studying such faint meteors with optical magnitudes of 10--14 \cite{2002mea..book..123B}.
Thus, carrying out a long-term survey of a population of faint
meteors, covering the entire range of velocities from 12.3 to 71.9 km/s, will allow us to study the
features of their optical emission, taking into account possible fragmentation, and compare
it with dynamic models of micrometeoroids. Another directions of meteor study might be the analysis of their mass distribution both in the showers and in the sporadic component, as well as massive colorimetry of events during the peaks of different meteor showers in order to compare their spectral properties.

UV Ceti type flare stars belong to the main sequence red dwarfs, making up 70-75\% of the population of the Galaxy \cite{2021A&A...650A.201R}. The duration of their stay on the main sequence exceeds the age of the Universe, and during this time their main physical characteristics remain unchanged \cite{2016ApJ...823..102C}. These features determine great interest in red dwarfs as possible regions where life arises \cite{2023ApJ...954L..50E}. The probabilities of the occurrence of Earth-like planets orbiting these stars in habitable zones with a characteristic size of about 0.2 AU \cite{1993Icar..101..108K}, estimated from various observations, are in the range of 30 -- 50\% \cite{2015ApJ...807...45D,2020MNRAS.498.2249H}. This gives about 100 habitable Earth-like planets in the vicinity of 10 pc, and several billion in the entire Galaxy \cite{2023ApJ...954L..50E}. 
The small masses and sizes, as well as the low luminosity of red dwarfs, as a result of which the habitable zones are located close to the star, determine the relatively high probability of detecting Earth-like planets during transits. Their depth for earth-like planets ranges from 0.002 to 0.0084 magnitudes with the radius of the host star from 0.6 to 0.1 solar, and the probability of a successful ``edge-on'' orientation of the orbit of the star-planet system for an observer is 0.5 -- 1\% \cite{Cameron2016}. The photometric accuracy required for transit detection is close to the limit for ground-based instruments. Nevertheless, it is achieved in various programs, thanks to the optimal strategy of observations and selection of objects, the use of effective methods of data analysis and methods of modeling the processes of obtaining them. Thus, when studying several dozen red dwarfs of 12--15 magnitudes  with systems of distributed telescopes with diameters of 0.2 to 2 meters at exposures of 30--120 seconds, the measurement accuracies in the range 0.001 -- 0.004 magnitudes were achieved, leading to the detection of effects with a depth of 0.005 -- 0.007 magnitude for planets of 1 -- 3 Earth masses located in habitable zones, with a probability of 2.5 - 8\% \cite{2012MNRAS.424.3101G, 2013ApJ...775...91B, 2020AJ....159..169G}. These results were obtained from large data sets using BLS \cite{2002A&A...391..369K, 2005MNRAS.356..557K} and TLS \cite{2019A&A...623A..39H} folding algorithms in the space of periods and filling factors.

In this context, it seems that such studies using the SAINT complex should be very effective. Instead of simultaneously monitoring up to 10 red dwarfs (according to the number of instruments used) within the framework of the mentioned programs, SAINT will be able to observe more than a thousand stars. With a minimum brightness of objects with potentially habitable Earth-like planets at a level of 16 magnitude, they will be located at a distance of up to 100 pc, their number in the northern sky will be about 100,000 \cite{2023ApJ...954L..50E}, and transits of these planets may be detected at several hundred red dwarfs. Note that when observing each star about 50 times over a visibility period of 100 days and a transit duty cycle of 0.001 -- 0.005 \cite{Cameron2016} the accuracy of intensity determination in phased light curves increases by tens of times relative to its initial estimates in individual exposures.  
Thus, as result of monitoring the sky with the SAINT complex, it is possible to increase the sample of Earth-like planets in the habitable zones by 4–6 times relative to its current volume \footnote{See Habitable Exoplanet Catalogue at \url{http://phl.upr.edu/projects/habitable-exoplanets-catalog}.}.

Similar results within the framework of the strategy for detecting periodic signals in wide-angle surveys can be obtained when searching for transits of Earth-like (and not only) planets around white dwarfs. These objects may also host planetary systems and Earth-like planets located in habitable zones. Indeed, with a luminosity of $10^{-3} - 10^{-4}$ and a mass of 0.5 -- 0.6 solar, their radii are close to those of the Earth, and the temperature lies in the range of 4000 -- 9000 K, which leads to the existence of a habitable zone  in the  0.005 - 0.02 AU range of distances from stars over several billion years \cite{1993Icar..101..108K, 2011ApJ...731L..31A, 2011tfa..confE..25A}. On the other hand, there are numerous evidences of the existence of various proto-planetary and post-planetary structures around white dwarfs -- disks, planetesimals, asteroids \cite{2017NatAs...1E..32F, 2021ApJ...922....4S}. Finally, 4 giant planets were discovered in systems with white dwarfs \cite{2021orel.bookE...1V}, one of them by transit method \cite{2020Natur.585..363V}. Thus, one can hope also for the presence of Earth-like (rocky) planets with a mass of 1 -- 2 of Earth one in the habitable zones  of white dwarfs, as established for 63 G-M dwarfs (according to Habitable Exoplanet Catalogue).  
The depth of their transits, repeated with periods in the interval of 4 -- 30 hours (corresponding to the size of the habitable zone), is 10 -- 100\%, the duration is 1 -- 2 minutes, and the probability of the best ``edge-on'' orientation of the orbital plane relative to the observer is close to 1 -- 2\% for planet radii 1 - 2 Earth's \cite{2011ApJ...731L..31A, 2011tfa..confE..25A, 2011MNRAS.410..899F}. In ground-based observations, even with small-diameter telescopes and standard photometric accuracy, it is possible to search for such effects. Primary difficulty in such studies is the needed relatively high temporal resolution at the level of several seconds. In particular, \cite{2011MNRAS.410..899F} searched for transits in long-term monitoring of 194 white dwarfs of 9 -- 15 mag with a time resolution of 30 seconds (an 8-channel SuperWASP telescope was used -- see Fig.~\ref{fig:exposure_limits} and Table~\ref{tab:surveys}) and obtained only upper limits at 10\% for the frequency of occurrence of planets and brown dwarfs. The authors come to the conclusion that it is necessary to increase the studied sample, the duration of observations, and to improve the temporal resolution. Similar programs have been planned, and are being partially implemented using space-borne instruments \cite{2013arXiv1309.0009K, 2021AAS...23713405G}, ground-based telescopes \cite{2019CoSka..49..380K, 2023RMxAC..55...93C} and their combinations \cite{2021tsc2.confE.119A}.
 
In the northern sky, according to various estimates, there are 10 -- 15 thousand white dwarfs brighter than 18 magnitude (see, for example, \cite{2013ApJS..204....5K} or \cite{2021MNRAS.508.3877G}), and for approximately 100 -- 150 ones the transits with periods between 3 hours and 50 days, with duty cycles of 0.03 -- 0.0006 and depths 0.1 -- 1, may be detected by SAINT. Note, however, that this estimate is solely for the objects where it is possible to detect planets, if they are there -- and the probability of it is unknown to us. On the other hand, in the absence of an effect, it will be possible to obtain an estimate of the upper limit for this probability.

Finally, there are several other (high risk) directions placed in the area of "Terra incognita" in the Fig.~\ref{fig:saint_limits} where SAINT observations may be potentially crucial:
\begin{enumerate}
\item detection of optical flashes accompanying gamma-ray bursts (about 5 events per year?),
\item registration of optical companions of fast radio bursts, in particular, the ones with repeating activity, where the same approach as for exoplanets may be used to improve the performance, 
\item registration of optical companions of gravitational wave events or setting  upper limits on their electromagnetic counterparts energy,
\item search for signals of terrestrial civilizations.
\end{enumerate}
 
\section{Conclusions}
\label{sec:conclusions}

In this paper, we present a project of a SAINT (Small Aperture Imaging Network Telescope) -- robotic 12-telescope wide-field complex with high temporal resolution, aimed towards detection and studying optical transient phenomena on shortest possible time scales. Its multi-channel nature allows different modes of operation depending on the task -- e.g. wide-field survey or a narrow-field follow-up. Using modern large-format CMOS detectors allows operation with the exposures as short as 20 milliseconds, and to observe the whole sky once every two nights. Novel reducer-collimator that may be rapidly installed into the light beam allows increasing the spatial resolution and effectively converting the complex to a photopolarimeter able to simultaneously acquire both color and polarimetric information.

Flexible nature of the complex, and its operation in the poorly studied region of parameter space for optical transients faster than a second, will provide an overwhelming amount of data on various classes of both astrophysical and near-Earth objects. On the other hand, as the complex is built from mostly off-the-shelf components, it may be easily scaled horizontally, to either increase the performance of an individual installation, or to expand it to several sites over the globe able to provide a continuous view of the entire sky, thus realizing the ``all sky all time'' concept which is extremely important when searching for any transient events, especially short ones.

\vspace{6pt} 



\section*{Author contributions}
Conceptualization, G.B., G.O., S.K. and A.G.; methodology, G.B., A.G..; software, S.K.; validation, G.V., A.V. and V.V. and N.L.; formal analysis, A.G., A.B.; investigation, N.L., V.S.; data curation, S.K., N.L.; writing---original draft preparation, G.B., G.O, A.G., S.K. and A.B.; visualization, A.G, A.B.; supervision, G.B., G.O.; project administration, G.B., G.O. and G.V.; funding acquisition, G.B., S.K., G.O. and G.V. All authors have read and agreed to the published version of the manuscript.

\section*{Funding}
The work was partially supported by the European Structural and Investment Fund and the Czech Ministry of Education, Youth and Sports (project CoGraDS CZ.02.1.01/0.0/0.0/15 003/0000437) and the Federal Program for Improving Competitiveness of the Kazan (Volga Region) Federal University.

\section*{Data availability}
The data presented in this study are available on request from the corresponding author.

\section*{Acknowledgments}
The work was carried out within the framework of the state assignment of the SAO RAS, approved by the Ministry of Science and Higher Education of the Russian Federation.
The authors are grateful to Lisa Chmyreva and Tatiana Sokolova for their help with the preparation of the manuscript.

\section*{Conflicts of interest}
The authors declare no conflict of interest.






\section*{APPENDIX: Current survey wide-field telescopes and systems}
In the table below main parameters of the current wide-field observational systems are presented. Remarks to the table content are as follows: \textbf{(a)} For multilens systems etendue $A\Omega$ have been calculated as sum of etendues of all individual lenses in the system. \textbf{(b)} The limit is given for the 8.5cm camera. \textbf{(c)} There is also another regime with $t_\mathrm{exp} = 30$ s and $m_\mathrm{lim} = 19.5$ \textbf{(d)} The limit is given in $m_{AB}$ and depend on the band chosen. \textbf{(e)} Exposure time varies from 30 to 45 seconds depend on the band chosen. Typically sum of 10 successive images is then used to achieve deeper limit. \textbf{(f)} The exposure time is given, while the total image cycle time is equal to 6 seconds.\textbf{(g)} Effective exposure time is given, since TDI is adopted in SDSS.\textbf{(h)} The project ``Pi of the Sky'' stopped gathering data in 2017 \cite{2022icrc.confE.536P} \textbf{(i)} Currently only one (ROTSE III-d) telescope is in operating. 

\begin{landscape}

\begin{table}
\caption{Summary of wide field survey telescopes currently at operation. The parameters of the SAINT are also included to the bottom of the table.\label{tab:surveys}}
\begin{tabular}{rlrrrrrl}
\hline
\textbf{N} & \textbf{Survey} & \textbf{$t_\mathrm{exp}$, $s$} & \textbf{FOV, deg$^2$} & \textbf{$m_\mathrm{lim}$} & \textbf{D, m} & \textbf{$A\Omega^a$, m$^2$deg$^2$} & \textbf{References} \\
\hline
1	& Evryscope												&   120		&	8150	&    16		&	22$\times$0.06		 & 23.0 & \cite{2019PASP..131g5001R}	\\
2	& Pi of the Sky											&	10		&	6400	&    12		&	16$\times$0.085    	 & 36.3 & \cite{2014RMxAC..45....7M}	\\
3	& Ground-based Wide Angle Cameras (GWAC)				&	10	  	&	5000	&    16		&	40$\times$0.18		 & 127  & \cite{2021PASP..133f5001H}	\\
4	& Arctic Wide-field Cameras (AWCam)						&   10		&	1799	&    12.6 	&	0.085 + 0.050	     & 5.40 & \cite{2013AJ....145...58L}	\\
5	& RAPTOR            									&  	30    	&  	1500    &	 12.5   &   8$\times$0.085		 & 8.51 & \cite{2002SPIE.4845..126V}	\\
6	& Kilodegree Extremely Little Telescope (KELT)			&   150		&	1352	&    12		&	2$\times$0.042		 & 1.87 & \cite{2012PASP..124..230P}	\\
7	& SuperWasp												&	30		&	1000	&	 15		&	16$\times$0.2		 & 31.4 & \cite{2006PASP..118.1407P}	\\
8	& Mini-MegaTORTORA (MMT-9)								&	0.1		&	950		&    12		&	9$\times$0.071		 & 3.76 & \cite{2017AstBu..72...81B}	\\
9   & Large Array Survey Telescope (LAST)					&	20		&	355		&	 19.6 	& 	48$\times$0.28		 & 2.96 & \cite{2023PASP..135f5001O, 2023arXiv230402719B}    \\
10	& Argus													&	1		&	344		&    16.5  	&	38$\times$0.203	     & 11.1 & \cite{2021AAS...23723502L}	\\
11	& HATSouth												&   240   	&	201     &    18.5   &   12$\times$0.18	     & 5.11 & \cite{2013PASP..125..154B}	\\
12	& ASAS-SN												&	90		&	108		&  	 17		&	24$\times$0.14		 & 1.66 & \cite{2017PASP..129j4502K}	\\
13	& The Next Generation Transit Survey (NGST) 			&	10		&	96		&	 16		&	12$\times$0.2		 & 3.02 & \cite{2018MNRAS.475.4476W}	\\
14	& MASTER-II												&	30		&	80		&	 20		&	20$\times$0.18		 & 2.04 & \cite{2012ExA....33..173K}	\\
15	& Deca-Degree Optical Transient Imager (DDOTI)			&	60		&	72		&    18		&	6$\times$0.28		 & 4.43 & \cite{2016SPIE.9910E..0GW}	\\
16	& Zwicky Transient Facility	(ZTF)						&	30		&	47		&    20		&	1.2			         & 53.2 & \cite{2019PASP..131a8002B}	\\
17	& GW Optical Transient Observatory (GOTO)				&	60		&	40		&    19		&	8$\times$0.40		 & 5.03 & \cite{2020SPIE11445E..7GD}	\\
18	& Tomo-e Gozen											&	0.5 	&	21		&  	 17		&	1.0			         & 16.5 & \cite{2018SPIE10702E..0JS} \\
19	& Catalina Sky Survey 0.7m								&	30		&	19.4	&    19.5	&	0.7			         & 7.47 & \cite{2003DPS....35.3604L, 2009ApJ...696..870D}	\\
20	& ATLAS													&	30		&	15		&    19		&	2$\times$0.5		 & 2.95 & \cite{2018PASP..130f4505T}	\\
21	& ROTSE III 											&	5		&	13.7	& 	 17		&	4$\times$0.45		 & 2.18 & \cite{2003PASP..115..132A}	\\
22	& Vera C. Rubin observatory (LSST)						&	30		&	9.6		&  	 25		&	6.5			         & 318  & \cite{2019ApJ...873..111I}	\\
23	& BlackGEM          									&   300   	&	8.1     &    22.2   &   3$\times$0.6		 & 2.29 & \cite{2022SPIE12182E..1VG}	\\
24	& La Silla-QUEST Variability survey 					&	60		&	7.5		&    20.5	&	1.0			         & 5.89 & \cite{2021AAS...23732403C}	\\
25	& Weizmann Fast Astronomical Survey Telescope (W-FAST)	&	0.04  	&	7		&	 13		&	0.55		         & 1.66 & \cite{2021PASP..133g5002N}	\\
26	& PanSTARRS-1											&	30		&	7		&    18.5	&	1.8			         & 17.8 & \cite{2016arXiv161205560C}	\\
27	& Space Surveillance Telescope (SST)					&	1		&	6		& 	 20.6	&	3.5			         & 57.7 & \cite{2014Icar..239..253R}	\\
28	& SkyMapper												&	40		&	5.5		&    18		&	1.35		         & 7.87 & \cite{2007PASA...24....1K}	\\
29	& Catalina Sky Survey 1.5m								&	30		&	5		&    21.5	&	1.5			         & 8.84 & \cite{2003DPS....35.3604L, 2009ApJ...696..870D}	\\
30	& Deep Lens Survey										&	600	 	& 	4		&    24		&	2$\times$4			 & 50.3 & \cite{2004ApJ...611..418B}	\\
31	& Sloan Digital Sky Survey	(SDSS)						&	54		&	3		&    22		&	2.5		             & 14.7 & \cite{2000AJ....120.1579Y}	\\
\hline
\textbf{32} & \textbf{SAINT} & \textbf{0.020} & \textbf{468} & \textbf{12.6} & \textbf{12$\times$0.3} & \textbf{33} & \textbf{This work} \\

\hline
\end{tabular}
\end{table}

\end{landscape}

\bibliographystyle{ieeetr-etal}
\bibliography{saint}

\end{document}